%% file: paper.tex
\documentclass[letterpaper,journal]{IEEEtran}
\usepackage{amsmath,amssymb,amsfonts,amsthm,mathtools}
\usepackage{graphicx,booktabs,array,cite}
\usepackage[caption=false,font=footnotesize]{subfig}
\usepackage{algorithm,algorithmic}
\usepackage{microtype,url,placeins,flushend}
\usepackage[hidelinks]{hyperref}
\graphicspath{{figures/}}

\newcommand{\R}{\mathbb{R}}

\newcommand{\x}{\mathbf{x}}
\newcommand{\F}{\mathbf{F}}
\newcommand{\W}{\mathbf{W}}
\newcommand{\Z}{\mathbf{Z}}
\newcommand{\I}{\mathbf{I}}
\newcommand{\G}{\mathbf{G}}
\newcommand{\J}{\mathbf{J}}
\newcommand{\norm}[1]{\left\lVert#1\right\rVert}

\DeclareMathOperator{\Rea}{Re}
\DeclareMathOperator{\Ima}{Im}
\allowdisplaybreaks

\hypersetup{pdftitle={Reduced-Order Model Characterization of Nonlinear Sustained Oscillations},pdfauthor={Kaiyang Huang, Dan Wilson, Kai Sun}}
\begin{document}
\title{Reduced-Order Model Characterization
of Nonlinear Sustained Oscillations}
\author{Kaiyang Huang,~\IEEEmembership{Member,~IEEE, }Dan Wilson, Kai Sun,~\IEEEmembership{Fellow,~IEEE. }
        % <-this % stops a space
\thanks{This work was supported by NSF grants ECCS-2329924 and CMMI-2140527.}
\thanks{K. Huang is with the Department of Electrical Engineering and Computer Science, South Dakota State University, Brookings, SD 57007 USA, and also with the Department of Electrical Engineering and Computer Science, University of Tennessee, Knoxville, TN 37996 USA (e-mail: kaiyang.huang@sdstate.edu).}%
\thanks{D. Wilson and K. Sun are with the Department of Electrical Engineering and Computer Science, University of Tennessee, Knoxville, TN 37996 USA (e-mail: dwilso81@utk.edu,kaisun@utk.edu).}}% <-this % stops a space}
\maketitle
\begin{abstract}
Nonlinear sustained oscillations involving inverter-based resources can happen when device and network interactions produce an attracting limit cycle around an unstable equilibrium. A damping controller must then describe nonlinear dynamics over the region between the two invariant sets, where a local equilibrium linearization can be insufficient. In this paper, a controlled phase--isostable model is first developed on a two-dimensional invariant manifold connecting the equilibrium and the limit cycle. Dynamics on this surface are represented by phase and isostable coordinates, while physical states and control inputs are related to these coordinates through nonlinear reconstruction and response functions. The required functions are computed from connecting trajectories and periodic-orbit data. A control strategy is then designed to provide extra damping while limiting phase distortion. Case studies on a two-area system and an IEEE 39-bus system demonstrate more effective oscillation damping than the compared equilibrium-linearized designs under the same conditions.
\end{abstract}
\input{sections/introduction}
\input{sections/model}
\input{sections/numerical}
\input{sections/control}
\input{sections/cases}

\FloatBarrier
\section{Conclusion}\label{sec:conclusion}
A two-state nonlinear model has been developed to describe sustained oscillations in power systems with an unstable equilibrium and an attracting limit cycle. Through this formulation, the effects of physical inputs on oscillation phase and inward motion toward the equilibrium are represented in reduced coordinates. Based on the nonlinear radial response, a damping control strategy has been constructed subject to actuator and phase-speed bounds. Case studies show that, stronger attenuation of physical oscillations has been achieved with lower input effort than with the controller based on the linearized model under the same actuator limits. The improvement has also been observed in unactuated devices, supporting the use of nonlinear input responses for system-wide oscillation damping. These results demonstrate the effectiveness of the proposed approach over the tested operating conditions.
\section*{Acknowledgments}
During manuscript preparation, ChatGPT (OpenAI) was used to assist with checking the mathematical derivations in Sections II–IV and with language editing. Responsibility for the final content is retained by the authors.
\appendices
\input{sections/appendix}
\bibliographystyle{IEEEtran}
\bibliography{references}
\end{document}

%% file: sections/introduction.tex
\section{Introduction}\label{sec:introduction}
\IEEEPARstart{S}{ustained} oscillations in power systems can involve nonlinear interactions among synchronous generators, inverter controls, and the network. Following a loss of equilibrium stability, these interactions may produce an attracting periodic orbit, so that the system continues to oscillate after the initiating disturbance has passed \cite{Abed1984,Tan1993}. In this regime, damping requires the applied control to drive the state away from the periodic orbit and toward the operating equilibrium. A model for this task must describe how the oscillation and its response to control change throughout that transition.

Several mechanisms can lead to such behavior in systems with inverter-based resources (IBRs). Interactions between a phase-locked loop (PLL) and current controllers can give rise to a stable limit cycle through a supercritical Hopf bifurcation \cite{Ma2020}. Current-controller limiters can also determine the amplitude and frequency of sustained oscillations, as studied using describing-function models in \cite{Huang2022DF}. More recent analysis has shown that slow load and voltage-control dynamics can induce oscillatory instability in faster inverter controls \cite{Ospina2025}. These findings motivate a description of finite-amplitude motion that retains the effects of both device and network dynamics.

Equilibrium-based methods provide an established foundation for oscillation analysis and damping design. Modal sensitivities and residues identify the influence of control inputs on linear modes \cite{Pagola1989}, while eigenvalue and singular-value indices help assess proximity to a Hopf bifurcation \cite{Canizares2004}. Nonlinear feedback has also been developed to relocate Hopf bifurcations and enlarge the region of stable operation \cite{Vahdati2017I,Vahdati2017II}. The present problem concerns control applied after a sustained oscillation has developed. During its suppression, both the timing and magnitude of the response to an actuator may vary with oscillation amplitude. An equilibrium linearization describes their local behavior, but does not generally retain this variation over the full transient. Direct use of a detailed nonlinear model is possible, although its many device states complicate the interpretation and design of the damping input.

Model reduction offers a way to retain the relevant nonlinear dynamics with fewer variables. Phase reduction describes the timing of an oscillation and its response to perturbations \cite{Nakao2016}. Damping also requires a coordinate for motion transverse to the periodic orbit. Isostable coordinates provide this information by assigning states according to the asymptotic decay of their transient components \cite{Mauroy2013,KVALHEIM2021132959,WilsonMoehlis2016}. Together, phase and isostable coordinates separate oscillatory motion from the decay toward the attracting cycle. Their response functions can be evaluated away from the cycle \cite{Shirasaka2017}, allowing the reduced description to account for changing input sensitivity.

Nonlinear phase-amplitude expansions and adaptive reductions extend this approach to larger departures from a reference oscillation \cite{Wilson2020,WilsonAdaptive,WilsonSun2024}. Related modeling has characterized nonlinear subsynchronous oscillations in a two-area power system containing an IBR \cite{Huang2025SSO}. For damping an established oscillation, the reduced model must cover the region extending inward from the limit cycle toward the equilibrium. The slow-manifold framework in \cite{WilsonSlow2026} provides a geometric basis for this extension: under suitable spectral and attraction conditions, trajectories leaving an unstable equilibrium and approaching a stable cycle form a low-dimensional invariant surface. Coordinates on this surface can describe the transition while a nonlinear reconstruction recovers the physical states.

The key contributions of this work are as follows. First, a controlled phase--isostable model is formulated on the two-dimensional invariant manifold connecting the unstable equilibrium and the attracting limit cycle. Finite-amplitude dynamics are retained through nonlinear state reconstruction and phase- and amplitude-dependent input responses. Second, a damping law is formulated from the nonlinear radial response, by which inward motion is promoted while phase distortion and actuator magnitude are constrained. In this way, the invariant geometry, its numerical construction, and its power-system control application are placed in one coordinate framework.

Section~\ref{sec:model} develops the reduced representation from the underlying geometry. Section~\ref{sec:numerical} describes its computation, and Section~\ref{sec:control} uses the input responses for damping design. Section~\ref{sec:cases} presents the two system studies, followed by conclusions in Section~\ref{sec:conclusion}.

%% file: sections/model.tex
\section{Slow-Manifold-Based Reduced Modeling}\label{sec:model}
In this section, a controlled phase--isostable model is formulated on a slow invariant manifold to represent nonlinear sustained oscillations in power systems. Physical actuator channels are incorporated through state-dependent input response vectors, and nonlinear state reconstruction is used to recover the physical states from the reduced coordinates. 

\subsection{Power system dynamics}
The dynamic behavior of a power system can be represented as nonlinear differential equations
\begin{equation}\label{eq:physical}
 \dot{\x}=\mathbf{f}(\x,\mathbf{u})
 =\F(\x)+\mathbf{B}(\x)\mathbf{u}
 +O(\norm{\mathbf{u}}_2^2),
\end{equation}
where $\x\in\R^n$ contains the dynamic device states, and $\mathbf{u}\in\R^p$ contains the supplementary control inputs. If the vector field is twice continuously differentiable in $\mathbf{u}$ near $\mathbf{0}$, its input dependence is written as $\mathbf{B}(\x)=\partial_{\mathbf{u}}\mathbf{f}(\x,\mathbf{0})\in\R^{n\times p}$ where $\F(\x)=\mathbf{f}(\x,\mathbf{0})$. Each column of $\mathbf{B}$ gives the differential state-space direction of one input channel after network elimination. For a control-affine model, the remainder in \eqref{eq:physical} vanishes. If the control parameters of some IBRs are not properly selected, the operating equilibrium may lose stability through an oscillatory mode, and the resulting nonlinear dynamics may approach a stable limit cycle \cite{Ma2020,Huang2022DF}. To describe this behavior, the autonomous model $\dot{\x}=\F(\x)$ is considered, with $\F\in C^3(\mathcal{D},\R^n)$ on an open domain $\mathcal{D}\subset\R^n$. Suppose that $\F(\x_e)=\mathbf{0}$ and let $\mathbf{A}=D\F(\x_e)$. The equilibrium is assumed to have one unstable complex-conjugate eigenvalue pair: $\sigma(\mathbf{A})=\{\lambda_u,\overline{\lambda}_u\}\cup\sigma_s$, where $\lambda_u=\alpha+\mathrm{i}\beta$, $\alpha>0$, $\beta\ne0$, and $\Rea\lambda<0$ for every $\lambda\in\sigma_s$. Suppose further that a hyperbolic attracting periodic orbit $\Gamma=\{\x_\Gamma(t):0\leq t<T_\Gamma\}$ exists, with $\dot{\x}_\Gamma=\F(\x_\Gamma)$ and $\x_\Gamma(t+T_\Gamma)=\x_\Gamma(t)$. The reduced model is developed for trajectories connecting $\x_e$ to $\Gamma$.

\subsection{Phase, isostables, and state reconstruction on slow manifolds}
The equilibrium and periodic orbit specify the limiting states of the connecting motion, and the intermediate state-space region is therefore characterized through their invariant manifolds. Let $\boldsymbol{\Phi}_t$ denote the autonomous flow, the unstable manifold of $\x_e$ is
\begin{equation}\label{eq:unstablemanifold}
 W^u(\x_e)=\left\{\x\in\mathcal{D}:
 \lim_{t\to-\infty}\boldsymbol{\Phi}_t(\x)=\x_e\right\}.
\end{equation}
Under the spectral assumption above, $W^u(\x_e)$ is
a two-dimensional, generally nonlinear manifold.
Near $\x_e$, a first-order approximation is provided
by the affine plane $    \x_e+\operatorname{span}_{\R}
    \{\Rea\mathbf{v}_u,\Ima\mathbf{v}_u\}$,
where $\mathbf{v}_u$ is a right eigenvector associated
with $\lambda_u$.
The backward trajectories in
\eqref{eq:unstablemanifold} are required to exist
for all $t\leq 0$.
Since $\Gamma$ is attracting, its basin of attraction
in $\mathcal{D}$ is denoted by $W^s(\Gamma)$. Suppose that
\begin{equation}\label{eq:manifold}
 \mathcal{M}=W^u(\x_e)\cap W^s(\Gamma)
\end{equation}
is nonempty, then $\mathcal{M}$ is a two-dimensional invariant manifold, with $\{\x_e\}\cup\Gamma\subset\overline{\mathcal{M}}$. Note transverse decay rate toward the manifold is governed by $\max_{\lambda\in\sigma_s}\Rea\lambda$
near $\x_e$ and by the slowest-decaying Floquet mode transverse
to $\mathcal{M}$ near $\Gamma$, with the local decay rates
varying continuously along the connecting trajectories
\cite{WilsonSlow2026}. Having identified the manifold, coordinates are needed to distinguish oscillatory motion from relaxation toward the limit cycle. These motions are described by a phase $\theta$ and a retained isostable coordinate $\psi$, respectively. Parameterize the cycle by $\x_\Gamma(\theta)$, with $\omega=2\pi/T_\Gamma$ and $\mathbb{S}^1=\R/(2\pi\mathbb{Z})$. The asymptotic phase $\theta:W^s(\Gamma)\to\mathbb{S}^1$ is defined by
\begin{equation}\label{eq:asymptoticphase}
 \lim_{t\to\infty}\norm{\boldsymbol{\Phi}_t(\x)
 -\x_\Gamma(\theta(\x)+\omega t)}_2=0.
\end{equation}
The selected real isostable coordinate $\psi$ is associated with a retained
real Floquet exponent $\kappa<0$, which describes radial
relaxation toward $\Gamma$ on the same invariant manifold
generated from the unstable eigenspace at $\x_e$. The coordinate functions are assumed to be $C^3$ in a neighborhood of the modeled manifold and to satisfy \cite{WilsonMoehlis2016,Shirasaka2017}
\begin{equation}\label{eq:coordinateflow}
 \begin{aligned}
 \theta(\boldsymbol{\Phi}_t(\x))
   &=\theta(\x)+\omega t\pmod{2\pi},\\
 \psi(\boldsymbol{\Phi}_t(\x))&=e^{\kappa t}\psi(\x).
 \end{aligned}
\end{equation}
The sign is chosen so that $\psi=0$ on $\Gamma$ and $\psi>0$ on the modeled inward branch of $\mathcal{M}$. The level sets of $\theta$ and $\psi$ are isochrons and isostables, respectively. Since $\dot\psi=\kappa\psi<0$ along an uncontrolled connecting trajectory, motion toward the equilibrium corresponds to increasing $\psi$. The physical states must subsequently be recovered from these coordinates to evaluate device responses. For this purpose, regular coordinates are assumed on the modeled portion of $\mathcal{M}$, and a reconstruction map is introduced as
\begin{equation}\label{eq:reconstruction}
 \x=\W_\psi(\theta,\psi),\qquad
 \operatorname{rank}D\W_\psi=2.
\end{equation}
Here the reconstruction $\W_\psi$ is supposed to be a local
diffeomorphism onto the modeled portion of $\mathcal{M}$,
with its local inverse given by the phase and isostable
coordinates. The reconstruction is then $2\pi$-periodic in $\theta$ and extends smoothly to $\W_\psi(\theta,0)=\x_\Gamma(\theta)$. Differentiating \eqref{eq:reconstruction} along \eqref{eq:coordinateflow} then gives the invariance equation
\begin{equation}\label{eq:invariancepsi}
 \omega\,\partial_\theta\W_\psi
 +\kappa\psi\,\partial_\psi\W_\psi=\F(\W_\psi).
\end{equation}
Consequently, autonomous physical trajectories on $\mathcal{M}$ are represented by \eqref{eq:coordinateflow} together with $\W_\psi$. 

\subsection{Input responses and the controlled reduced model}
While the preceding construction describes autonomous dynamics on the manifold, to incorporate the physical control inputs, their effects on phase and isostable coordinates must next be determined. The required response vectors are defined in the full state
space as $\Z=\nabla_{\x}\theta\in\R^n$ and
$\I=\nabla_{\x}\psi\in\R^n$. Differentiating \eqref{eq:coordinateflow} with respect to
time and applying the chain rule gives
\begin{equation}\label{eq:transport}
 \begin{aligned}
 \dot
 \theta(\boldsymbol{\Phi}_t(\x))
 &=\Z^{\mathrm T}\F=\omega,\\
\dot
 \psi(\boldsymbol{\Phi}_t(\x))
 &=\I^{\mathrm T}\F=\kappa\psi.
 \end{aligned}
\end{equation}
Since $\Z$ and $\I$ are gradients of scalar functions,
their Jacobians $D_{\x}\Z$ and $D_{\x}\I$ are symmetric.
Thus, with $\J=D_{\x}\F$, taking the spatial gradients
of \eqref{eq:transport} using the product rule gives
\begin{equation}\label{eq:differentiatedtransport}
 \begin{aligned}
 (D_{\x}\Z)\F+\J^{\mathrm T}\Z&=\mathbf{0},\\
 (D_{\x}\I)\F+\J^{\mathrm T}\I&=\kappa\I.
 \end{aligned}
\end{equation}
To obtain the time evolution of the response vectors,
the chain rule is then applied along an uncontrolled
trajectory $\dot{\x}=\F(\x)$, and substitution of
\eqref{eq:differentiatedtransport} yields the adjoint
equations \cite{WilsonSlow2026}:
\begin{equation}\label{eq:cycleadjoint}
 \begin{aligned}
 \dot{\Z}
 &=(D_{\x}\Z)\dot{\x}
  =(D_{\x}\Z)\F
  =-\J^{\mathrm T}\Z,\\
 \dot{\I}
 &=(D_{\x}\I)\dot{\x}
  =(D_{\x}\I)\F
  =(\kappa\mathbf{I}_n-\J^{\mathrm T})\I.
 \end{aligned}
\end{equation}
On $\Gamma$, these equations admit periodic solutions $\Z_\Gamma$ and $\I_\Gamma$. The phase scale is fixed by $\Z_\Gamma^{\mathrm T}\F=\omega$. After a scale and inward orientation have been chosen for the retained right Floquet mode $\mathbf{p}_r(\theta)$, the isostable scale is fixed by $\I_\Gamma^{\mathrm T}\mathbf{p}_r=1$; moreover, $\I_\Gamma^{\mathrm T}\F=0$. Thus, a periodic orbit provides the boundary values from which the response vectors can be continued along connecting trajectories. With these gradients available, the response to a general physical input follows directly from the chain rule applied to \eqref{eq:physical}:
\begin{equation}\label{eq:forcedcoordinates}
 \begin{aligned}
 \dot\theta&=\omega+\Z^{\mathrm T}\mathbf{B}\mathbf{u}
             +O(\norm{\mathbf{u}}_2^2),\\
 \dot\psi&=\kappa\psi+\I^{\mathrm T}\mathbf{B}\mathbf{u}
             +O(\norm{\mathbf{u}}_2^2).
 \end{aligned}
\end{equation}
Here the gradients and input matrix are evaluated at the physical state. To obtain a closed two-state model, they will subsequently be evaluated at the reconstructed state on $\mathcal{M}$. Before making this restriction, a logarithmic radial coordinate is introduced to accommodate the growth of $\psi$ toward the equilibrium. With a fixed dimensionless isostable normalization, define
\begin{equation}\label{eq:logcoordinate}
 \eta=\log(1+\psi),\qquad \G=\nabla\eta=e^{-\eta}\I.
\end{equation}
Thus, $\eta=0$ on $\Gamma$, and increasing $\eta$ retains the same inward orientation as increasing $\psi$. Substitution into \eqref{eq:forcedcoordinates} gives
\begin{equation}\label{eq:logdynamics}
 \dot\eta=\underbrace{\kappa(1-e^{-\eta})}_{a(\eta)}
 +\G^{\mathrm T}\mathbf{B}\mathbf{u}+O(\norm{\mathbf{u}}_2^2).
\end{equation}
The corresponding response evolution is obtained without introducing a separate coordinate approximation. Indeed, along an uncontrolled trajectory, differentiation of $\G=e^{-\eta}\I$ and substitution of \eqref{eq:cycleadjoint} give $\dot\G=e^{-\eta}\dot\I-a(\eta)\G$. Therefore, the equations used to propagate phase and logarithmic-isostable responses are
\begin{equation}\label{eq:zgadjoint}
 \dot\Z=-\J^{\mathrm T}\Z,\qquad
 \dot\G=(\kappa e^{-\eta}\mathbf{I}_n-\J^{\mathrm T})\G.
\end{equation}
The reduced model can now be closed by defining $\W(\theta,\eta)=\W_\psi(\theta,e^\eta-1)$ and evaluating the input responses on this reconstructed manifold:
\begin{equation}\label{eq:zg}
 \begin{aligned}
 \mathbf{z}(\theta,\eta)&=\mathbf{B}(\W)^{\mathrm T}\Z(\W)\in\R^p,\\
 \mathbf{g}(\theta,\eta)&=\mathbf{B}(\W)^{\mathrm T}\G(\W)\in\R^p.
 \end{aligned}
\end{equation}
Each entry specifies the response to one input channel. Retaining the first-order input terms then yields
\begin{equation}\label{eq:rom}
 \begin{aligned}
 \dot\theta&=\omega+\mathbf{z}(\theta,\eta)^{\mathrm T}\mathbf{u},\\
 \dot\eta&=a(\eta)+\mathbf{g}(\theta,\eta)^{\mathrm T}\mathbf{u},\\
 \widehat{\x}&=\W(\theta,\eta).
 \end{aligned}
\end{equation}
For $\eta>0$, the autonomous drift satisfies $a(\eta)<0$; hence, positive inward motion is induced when $\mathbf{g}^{\mathrm T}\mathbf{u}>-a(\eta)$. The approximation in \eqref{eq:rom} consists of truncating higher-order input terms and evaluating the responses on the retained manifold instead of the generally off-manifold controlled state.

%% file: sections/numerical.tex
\section{Numerical Construction of the Reduced Model}\label{sec:numerical}
In this section, the reconstruction $\W$ and response vectors $\mathbf{z}$ and $\mathbf{g}$ are constructed from trajectories. Periodic-orbit information is first corrected at finite trajectory endpoints and then propagated over the modeled manifold. Alignment on a common radial grid subsequently provides the continuous functions required by \eqref{eq:rom}.

\subsection{Connecting trajectories and periodic-orbit states}
The manifold is sampled by trajectories leaving the neighborhood of $\x_e$ along its unstable eigenspace. With the eigenvector convention of Section~\ref{sec:model}, the initial states are chosen as
\begin{equation}\label{eq:seeds}
 \x_k(0)=\x_e+2\Rea\{\varepsilon\mathbf{v}_u e^{\mathrm{i}\chi_k}\},
 \qquad \chi_k=2\pi k/N_s,
\end{equation}
for $k=0,\ldots,N_s-1$. These seeds approximate $W^u(\x_e)$ to leading order, with $\varepsilon>0$ controlling the initialization error. Forward integration follows the nonlinear connecting motion while strongly contracting transverse components decay. The state trajectories are stored for subsequent response calculations. The attracting cycle approached by the trajectories is refined by shooting, using $\boldsymbol{\Phi}_{T_\Gamma}(\x_\Gamma)=\x_\Gamma$ together with a scalar phase condition. Along this orbit, the variational equation $\dot{\mathbf{P}}=\J_\Gamma\mathbf{P}$, initialized by $\mathbf{P}(0)=\mathbf{I}_n$, gives the monodromy matrix $\mathbf{P}(T_\Gamma)$. Its neutral multiplier identifies the phase direction, while the retained positive stable multiplier $\mu_r$ gives $\kappa=\log(\mu_r)/T_\Gamma$. The retained mode is selected to represent the radial direction of the modeled manifold. The corresponding left Floquet eigenvectors provide initial values for the periodic solutions of \eqref{eq:cycleadjoint}. Their normalization is imposed as specified in Section~\ref{sec:model}, consistently with the selected right Floquet mode. Once $\Z_\Gamma$ and $\I_\Gamma$ have been obtained, the periodic Hessians $\mathbf{H}_\theta=\nabla^2\theta|_\Gamma$ and $\mathbf{H}_\psi=\nabla^2\psi|_\Gamma$ are computed from the equations in Appendix~\ref{app:hessian} \cite{Wilson2019}. These periodic quantities are defined on $\Gamma$, whereas each trajectory ends at a nearby state $\x_T$. The off-cycle terminal states are therefore obtained by a local expansion about $\x_\Gamma(\theta_T)$. Writing $\boldsymbol{\delta}=\x_T-\x_\Gamma(\theta_T)$, the terminal phase is estimated by solving
\begin{equation}\label{eq:terminalphase}
 \Z_\Gamma(\theta_T)^{\mathrm T}\boldsymbol{\delta}
 +\tfrac12\boldsymbol{\delta}^{\mathrm T}
 \mathbf{H}_\theta(\theta_T)\boldsymbol{\delta}=0.
\end{equation}
This equation enables the second-order phase approximation to assign the same phase to $\x_T$ and its reference point on the cycle. The same expansion supplies the isostable label and coordinate gradients:
\begin{equation}\label{eq:terminalcorrection}
 \begin{aligned}
 \psi_T&\approx\I_\Gamma^{\mathrm T}\boldsymbol{\delta}
       +\tfrac12\boldsymbol{\delta}^{\mathrm T}\mathbf{H}_\psi\boldsymbol{\delta},\\
 \Z_T&\approx\Z_\Gamma+\mathbf{H}_\theta\boldsymbol{\delta},\\
 \I_T&\approx\I_\Gamma+\mathbf{H}_\psi\boldsymbol{\delta}.
 \end{aligned}
\end{equation}
All periodic quantities are evaluated at $\theta_T$. For the assumed $C^3$ coordinates, the scalar and gradient truncation errors are $O(\norm{\boldsymbol{\delta}}_2^3)$ and $O(\norm{\boldsymbol{\delta}}_2^2)$, respectively. Before backward propagation, the approximate gradients are made consistent with the transport identities in \eqref{eq:transport}. With $\F_T=\F(\x_T)$, the corrected terminal gradients are set to
\begin{equation}\label{eq:terminalnormalize}
 \begin{aligned}
 \overline\Z_T&=\frac{\omega\Z_T}{\Z_T^{\mathrm T}\F_T},\\
 \overline\I_T&=\I_T+
 \frac{\kappa\psi_T-\I_T^{\mathrm T}\F_T}{\omega}\overline\Z_T.
 \end{aligned}
\end{equation}
The denominator is nonzero for sufficiently accurate states near $\Gamma$. Substitution verifies $\overline\Z_T^{\mathrm T}\F_T=\omega$ and $\overline\I_T^{\mathrm T}\F_T=\kappa\psi_T$. The states required for logarithmic-isostable propagation are then $\eta_T=\log(1+\psi_T)$ and $\G_T=\overline\I_T/(1+\psi_T)$, with $\psi_T>0$ on the selected inward branch.

\subsection{Trajectory labels and backward responses}
Once the terminal coordinates have been estimated, the coordinates
along each stored trajectory can be recovered using the known
autonomous evolution in \eqref{eq:coordinateflow}. Specifically,
for trajectory $k$ ending at time $T_k$, these flow relations are
inverted to obtain the initial phase and isostable labels:
\begin{equation}\label{eq:labels}
 \begin{aligned}
 \theta_{0,k}&=\theta_{T,k}-\omega T_k\pmod{2\pi},\\
 \psi_{0,k}&=e^{-\kappa T_k}\psi_{T,k}.
 \end{aligned}
\end{equation}
It follows that $\theta_k(t)=\theta_{0,k}+\omega t\pmod{2\pi}$ and $\eta_k(t)=\log(1+\psi_{0,k}e^{\kappa t})$. However, these labels inherit the finite-endpoint approximation error. Its sensitivity to the terminal time is checked by repeating the estimates at $T_k$ and $T_k-\Delta T$. Defining the resulting initial labels by superscripts gives
\begin{equation}\label{eq:labeldifferences}
 \begin{aligned}
 \Delta_{\theta,k}&=\operatorname{Arg}\exp\!\left[\mathrm{i}
 (\theta_{0,k}^{(T_k)}-\theta_{0,k}^{(T_k-\Delta T)})\right],\\
 \Delta_{\psi,k}&=\psi_{0,k}^{(T_k)}-\psi_{0,k}^{(T_k-\Delta T)}.
 \end{aligned}
\end{equation}
Phase and relative radial tolerances are applied to these differences before the trajectory is used for interpolation. Endpoint consistency is assessed through this comparison, provided that the local expansion is valid at both endpoints.With $\theta_k(t)$ and $\eta_k(t)$ determined, the response vectors
$\Z_k(t)$ and $\G_k(t)$ are computed by integrating
\eqref{eq:zgadjoint} backward from $T_k$ to $0$, subject to the
terminal conditions $\Z_k(T_k)=\overline\Z_{T,k}$ and
$\G_k(T_k)=\G_{T,k}$. During this integration, the Jacobian
$\J(t)=D\F(\x_k(t))$ is evaluated along the previously computed
forward trajectory $\x_k(t)$.
To construct $\W$, $\mathbf{z}$, and $\mathbf{g}$ as functions of
$(\theta,\eta)$, the previously computed state trajectories and response
vectors are evaluated at common values $\eta_j>0$ of the logarithmic
isostable coordinate. For trajectory $k$, the corresponding time is
determined by imposing $\eta_k(t_{k,j})=\eta_j$. Substitution of
$\eta_k(t)=\log(1+\psi_{0,k}e^{\kappa t})$ then gives
\begin{equation}\label{eq:sampletime}
 t_{k,j}=\frac{1}{\kappa}
 \log\!\left(\frac{e^{\eta_j}-1}{\psi_{0,k}}\right).
\end{equation}
The admissible values of $\eta_j$ are restricted to those satisfying
$0\leq t_{k,j}\leq T_k$ for every retained trajectory. With
$\theta_{k,j}=\theta_k(t_{k,j})$, the reconstruction and input responses
at these coordinates are evaluated as
\begin{equation}\label{eq:multiinputsamples}
 \begin{aligned}
 \W(\theta_{k,j},\eta_j)&=\x_k(t_{k,j}),\\
 \mathbf{z}(\theta_{k,j},\eta_j)
   &=\mathbf{B}(\x_k(t_{k,j}))^{\mathrm T}\Z_k(t_{k,j}),\\
 \mathbf{g}(\theta_{k,j},\eta_j)
   &=\mathbf{B}(\x_k(t_{k,j}))^{\mathrm T}\G_k(t_{k,j}).
 \end{aligned}
\end{equation}
Thus, the responses of all $p$ input channels are obtained at the same
coordinates as the reconstructed physical state. At $\eta=0$, these
quantities are evaluated directly on $\Gamma$, using
$\W(\theta,0)=\x_\Gamma(\theta)$ and the periodic response vectors
$\Z_\Gamma$ and $\G_\Gamma=\I_\Gamma$.

%The phase assigned to each evaluation point also depends on the
%estimated initial isostable coordinate. Specifically, at fixed
%$\eta_j$ and $\theta_{0,k}$, differentiation of the unwrapped phase
%$\theta_{k,j}=\theta_{0,k}+\omega t_{k,j}$ gives
%$\partial\theta_{k,j}/\partial\log\psi_{0,k}=-\omega/\kappa$.
%Consequently, relative errors in $\psi_{0,k}$ are amplified in the
%phase used for interpolation when $|\omega/\kappa|$ is large.
\subsection{Interpolation and model assembly}
The reconstruction and input responses obtained at the common radial
coordinates must be extended to continuous functions of $(\theta,\eta)$
so that the reduced dynamics can be evaluated between the computed
trajectories. At each $\eta_j$, the values of $\W$, $\mathbf{z}$, and
$\mathbf{g}$ in \eqref{eq:multiinputsamples} are therefore ordered by
asymptotic phase and periodically interpolated onto a common phase
grid. These functions are represented jointly by
$\mathbf{p}=[\W^{\mathrm T},\mathbf{z}^{\mathrm T},
\mathbf{g}^{\mathrm T}]^{\mathrm T}\in\R^{n+2p}$ and approximated
using a Fourier expansion in phase with spline coefficients in $\eta$:
\begin{equation}\label{eq:fourier}
 \mathbf{p}(\theta,\eta)\approx
 \Rea\!\left\{\sum_{\ell=-M}^{M}\mathbf{c}_\ell(\eta)
 e^{\mathrm{i}\ell[\theta+s(\eta)]}\right\}.
\end{equation}
To reduce the radial variation of the Fourier coefficients, a smooth
phase shift $s(\eta)$ is extracted from the first harmonic of a selected
oscillatory state component. This shift is incorporated only into the
approximation basis, so $\theta$ remains the asymptotic phase used
in \eqref{eq:rom}.

Once the continuous reconstruction has been obtained, its derivatives
are evaluated to verify that the reconstructed surface remains regular.
Accordingly,
$\mathbf{W}_q=[\partial_\theta\W,\partial_\eta\W]$
is evaluated by differentiating the reconstruction components of
\eqref{eq:fourier}. In particular, differentiation with respect to
$\eta$ includes both the spline derivatives and the contribution
$\mathrm{i}\ell s'(\eta)\mathbf{c}_\ell(\eta)$ from the phase shift.
The condition $\operatorname{rank}\mathbf{W}_q=2$ is then checked
throughout the modeled region to ensure that a two-dimensional
tangent space is retained.

The complete construction is summarized in
Algorithm~\ref{alg:construction}.

\begin{algorithm}[t]
\caption{Construction of the multi-input slow-manifold model}
\label{alg:construction}
\begin{algorithmic}[1]
\let\OriginalSTATE\STATE
\renewcommand{\STATE}{\par\OriginalSTATE}
\REQUIRE $\F$, $\mathbf{B}$;
seed parameters $N_s,\varepsilon$; phase and radial grids;
Fourier order $M$.
\ENSURE Reconstruction $\W$, input response vectors
$\mathbf{z},\mathbf{g}$.
\STATE Compute $\x_e$, $\Gamma$, the unstable eigenpair,
and the retained Floquet exponent $\kappa$.
\STATE Compute the periodic coordinate gradients using
\eqref{eq:cycleadjoint} and the periodic Hessians using
Appendix~\ref{app:hessian}.
\FOR{each initial condition in \eqref{eq:seeds}}
\STATE Integrate and store the uncontrolled connecting trajectory.
\STATE Estimate the terminal coordinates and normalize the
terminal gradients using
\eqref{eq:terminalphase}--\eqref{eq:terminalnormalize}.
\STATE Recover the initial phase and isostable coordinates using
\eqref{eq:labels}, and check endpoint consistency using
\eqref{eq:labeldifferences}.
\STATE Integrate \eqref{eq:zgadjoint} backward along the stored
state trajectory.
\STATE Determine the evaluation times using \eqref{eq:sampletime}
and compute $\W,\mathbf{z},\mathbf{g}$ at the common radial
coordinates using \eqref{eq:multiinputsamples}.
\ENDFOR
\STATE Include $\W,\mathbf{z},\mathbf{g}$ on $\Gamma$,
interpolate onto the common phase grid, and fit \eqref{eq:fourier}.
\STATE Differentiate the fitted $\W$ to obtain $\mathbf{W}_q$,
and check $\operatorname{rank}\mathbf{W}_q=2$ over the modeled region.
\end{algorithmic}
\end{algorithm}

%% file: sections/control.tex
\section{Damping Control Based on the Reduced Model}\label{sec:control}
In this section, the proposed phase--isostable representation is converted into a power-system damping law constructed from the nonlinear radial response. Actuator and phase-speed constraints are imposed through the input limit and the phase response. 
\subsection{Admissible input and inward motion}
While the reduced model in Section~\ref{sec:model} remains valid for $p$ inputs, here a single-input case $p=1$ used in the case studies momentarily is considered. Accordingly, the input and response functions
are written as $u$, $z$, and $g$, and the input matrix
$\mathbf{B}(\x)$ reduces to a single direction $\mathbf{b}(\x)$.

Since increasing $\eta$ corresponds to dynamics toward the
equilibrium, the control input is required to counteract the
autonomous drift $a(\eta)<0$ for $\eta>0$. The available input
is restricted by both actuator capacity and the allowable
change in phase speed. Specifically, the bounds $|u|\leq U$
and $|zu|\leq c_\theta\omega$ are imposed, where
$0<c_\theta<1$. Their intersection gives the admissible
interval $[-b_c,b_c]$, with
\begin{equation}\label{eq:bound}
 b_c=\begin{cases}
 \min\{U,c_\theta\omega/|z|\},&z\neq0,\\
 U,&z=0.
 \end{cases}
\end{equation}
By the phase equation in \eqref{eq:rom}, these bounds ensure
$(1-c_\theta)\omega\leq\dot\theta\leq(1+c_\theta)\omega$,
so the phase continues to advance while its deviation from
the autonomous rate is limited.

With the admissible interval determined, the feasibility of
inward motion can be assessed from the radial equation
$\dot\eta=a+gu$. At each $(\theta,\eta)$, the largest radial
rate permitted by \eqref{eq:bound} is
\begin{equation}\label{eq:vmax}
 v_{\max}=\max_{|u|\leq b_c}(a+gu)=a+|g|b_c.
\end{equation}
For $g\neq0$, this rate is attained by
$u=b_c\operatorname{sign}(g)$, which identifies the input
direction that promotes inward motion. Thus, an admissible
input producing $\dot\eta>0$ exists at the given coordinates
if and only if $v_{\max}>0$.
\subsection{Continuous bounded control for isostable ascent}
The preceding analysis identifies $\operatorname{sign}(g)$ as the
input direction that increases the radial rate. However, direct use
of this sign function produces discontinuous switching at zeros of
$g$. To retain the same direction with a continuous transition, a
regularized control law is constructed from the phase and radial
responses.

Let $M_z(\eta)$ and $M_g(\eta)$ be interpolated envelopes of $|z|$
and $|g|$ over phase. To account for local values that may exceed
these interpolated envelopes, define
$\widetilde M_z=\max\{M_z,|z|,\epsilon_r\}$ and
$\widetilde M_g=\max\{M_g,|g|,\epsilon_r\}$, where
$\epsilon_r>0$ prevents division by zero. The normalized radial
response $d_g=g/\widetilde M_g$ is then used to define the
bounded damping law
\begin{equation}\label{eq:smoothlaw}
 u_N=A_u\frac{d_g}{\sqrt{d_g^2+\delta^2}},\qquad
 A_u=\min\{U,c_\theta\omega/\widetilde M_z\},
\end{equation}
where $\delta>0$ regularizes the sign transition. The amplitude
$A_u$ is selected to satisfy the actuator and phase-speed
constraints, while the regularized factor determines the input
direction and reduces its magnitude near zeros of $g$.

Since $\widetilde M_z\geq|z|$ and
$|d_g|/\sqrt{d_g^2+\delta^2}\leq1$, the admissible bound
in \eqref{eq:bound} is respected. Moreover, multiplication
of \eqref{eq:smoothlaw} by $g$ gives
\begin{equation}\label{eq:smoothbounds}
 \begin{aligned}
 |u_N|&\leq A_u\leq b_c,\\
 gu_N&=\frac{A_u g^2}
 {\widetilde M_g\sqrt{(g/\widetilde M_g)^2+\delta^2}}\geq0.
 \end{aligned}
\end{equation}
Thus, the input contribution always opposes the negative
autonomous drift, yielding $\dot\eta=a+gu_N\geq a$.
Inward motion requires the stronger condition $gu_N>-a$.
For fixed nonzero $g$, the control input approaches
$A_u\operatorname{sign}(g)$ as $\delta\to0$, whereas increasing
$\delta$ reduces the input magnitude and broadens the transition
around $g=0$.

Although the instantaneous radial rate need not remain positive,
the phase-speed bound ensures
$\dot\theta\geq(1-c_\theta)\omega>0$.
Consequently, inward progress can also be evaluated over successive
phase revolutions: for an interval $[t_1,t_2]$ during which the
unwrapped phase advances by $2\pi$, net inward motion occurs when
$\eta(t_2)-\eta(t_1)=\int_{t_1}^{t_2}(a+gu_N)\,\mathrm{d}t>0$.
\subsection{Equilibrium-linearized control}
To assess the benefit of retaining the nonlinear input responses,
a comparison controller is constructed from the local dynamics
at $\x_e$ using the same actuator and input bound.
With $\Delta\x=\x-\x_e$, linearization gives
$\Delta\dot\x=\mathbf{A}\Delta\x+\mathbf{b}_e u$, where
$\mathbf{b}_e=\mathbf{b}(\x_e)$.
The unstable mode is extracted using a complex left eigenvector
satisfying
$\mathbf{w}_u^{\mathrm T}\mathbf{A}
=\lambda_u\mathbf{w}_u^{\mathrm T}$ and
$\mathbf{w}_u^{\mathrm T}\mathbf{v}_u=1$.
Here $\mathrm T$ denotes ordinary transpose, and an overbar
denotes complex conjugation.

Defining the modal coordinate $q=\mathbf{w}_u^{\mathrm T}\Delta\x$
and input coefficient $b=\mathbf{w}_u^{\mathrm T}\mathbf{b}_e$,
the projected dynamics are obtained as
$\dot q=(\alpha+\mathrm{i}\beta)q+bu$.
The influence of the input on the modal magnitude is then
determined by
\begin{equation}\label{eq:modalenergy}
 \frac{\mathrm d|q|^2}{\mathrm dt}
 =2\alpha|q|^2+2\Rea(\overline bq)u.
\end{equation}
Thus, a nonpositive input contribution is obtained by choosing
$u$ with the opposite sign to $\Rea(\overline bq)$.
For $b\ne0$, the corresponding saturated control law is defined as
\begin{equation}\label{eq:linearcontrol}
    u_L=\max\!\left\{-U,\,
        \min\!\left\{U,-2k\operatorname{Re}(\bar{b}q)\right\}
    \right\},
    \quad
    k=\frac{\alpha+\gamma}{|b|^2},
\end{equation}
where $\gamma>0$ is a prescribed damping parameter.
Saturation enforces the actuator bound while preserving the
sign of the damping contribution in \eqref{eq:modalenergy}.

The role of $\gamma$ can be established from the unsaturated closed-loop dynamics of the linear design model when saturation is inactive. Under the coordinate
rotation $\zeta=(\overline b/|b|)q$, the input coefficient becomes
real, and the dynamics of
$[\Rea\zeta,\Ima\zeta]^{\mathrm T}$ are governed by
\begin{equation}\label{eq:linearclosedmatrix}
 \mathbf{A}_{\mathrm{cl}}=
 \begin{bmatrix}
 \alpha-2k|b|^2&-\beta\\
 \beta&\alpha
 \end{bmatrix}
 =
 \begin{bmatrix}
 -\alpha-2\gamma&-\beta\\
 \beta&\alpha
 \end{bmatrix}.
\end{equation}
The resulting eigenvalues are
\begin{equation}\label{eq:linearpoles}
 -\gamma\pm\mathrm{i}\sqrt{\beta^2-(\alpha+\gamma)^2},
\end{equation}
provided $|\beta|>\alpha+\gamma$.
Under this condition, the unstable modal pair is shifted to
real part $-\gamma$, so $\gamma$ specifies its decay rate
in the unsaturated linear model. The performance of the
bounded input in the nonlinear physical system is evaluated
in the case studies.

%% file: sections/cases.tex
\section{Case Studies}\label{sec:cases}
The proposed reduced model and damping law are evaluated on a two-area system and a modified IEEE 39-bus system. The two-area study illustrates how the nonlinear input responses support damping over the region between the limit cycle and the equilibrium. The IEEE 39-bus study then examines whether a two-state control design remains effective when detailed device dynamics are retained in the physical system. In both studies, performance is assessed from the full-order nonlinear responses.

\subsection{Systems and performance measures}\label{sec:metrics}\label{sec:execution}
The two systems contain one and five grid-following (GFL) inverters, respectively. The principal operating and control settings are summarized in Table~\ref{tab:config}. The smooth law \eqref{eq:smoothlaw} is compared with the equilibrium-linearized law \eqref{eq:linearcontrol}; the tested linear gains satisfy $|\beta|>\alpha+\gamma$.
For each comparison, both designs are initialized at the same point $\x_0\in\Gamma$ at phase zero and are applied through the same GFL direct-axis current reference, corresponding to $p=1$. Each control waveform is generated by integrating its design model and is subsequently applied to the full-order nonlinear system. The input bound and control horizon are identical for the two designs. Unless otherwise stated, $c_\theta=0.4$, $\delta=0.15$, and $\gamma=1$ are used. Blue and red curves denote the physical responses under the linear and nonlinear designs, respectively. To compare deviations across physical states with different units, the normalized distance from the equilibrium is defined as
$d(t)=\norm{\mathbf{D}^{-1}[\x(t)-\x_e]}_2/d_\Gamma$,$d_\Gamma=\frac{1}{N_\Gamma}\sum_{j=1}^{N_\Gamma}
 \norm{\mathbf{D}^{-1}[\x_\Gamma(\theta_j)-\x_e]}_2$, where $\mathbf{D}$ is a positive diagonal matrix whose entries are determined from the uncontrolled limit cycle as $D_{ii}=\max\{\max_\theta x_{\Gamma,i}(\theta)-\min_\theta x_{\Gamma,i}(\theta),d_{i,\min}\}$. For both systems, $d_{i,\min}=0.01$ in the corresponding state's unit is imposed to limit the weight assigned to states with small cycle variations. The phases $\theta_j$ are uniformly spaced over one cycle; thus, $d_\Gamma$ represents the mean scaled distance of the uncontrolled cycle from the equilibrium. Performance is measured by the terminal distance $d(T_c)$, the minimum distance, and the first-entry time $t_{5\%}=\inf\{t\in[0,T_c]:d(t)\leq0.05\}$. The 5\% threshold is used only for performance evaluation. Control effort is measured by $E_u=\int_0^{T_c}u^2\,\mathrm dt$. The same state scales and cycle normalization are used for all compared designs within each system.

\begin{table}[t]
\caption{Model and Control Settings}\label{tab:config}
\centering\small\setlength{\tabcolsep}{4pt}
\begin{tabular}{@{}lcc@{}}\toprule
Quantity & Two-area & IEEE 39-bus\\\midrule
Physical / reduced states & $8/2$ & $74/2$\\
Cycle period $T_\Gamma$ (s) & $0.193370$ & $0.215740$\\
Radial exponent $\kappa$ ($\mathrm{s}^{-1}$) & $-0.11068$ & $-0.07666$\\
PLL gains $(K_p,K_i)$ & $(0.60536,1400)$ & $(1.8,1400)$\\
Voltage/current gain $K_q$ & $200$ & $2000$\\
Control horizon $T_c$ (s) & $10$ & $20$\\

Smoothing parameter $\delta$ & $0.15$ & $0.15$\\
\bottomrule\end{tabular}
\end{table}

\subsection{Two-area system}\label{sec:twoarea}
The two-area system is used to examine how the reconstructed manifold and input responses describe the suppression of an established oscillation. The 11-bus network contains three synchronous generators and one GFL inverter, with device models given in \cite{Huang2025SSO}. With the PLL gains in Table~\ref{tab:config}, the equilibrium has one unstable pair $0.05\pm33.1500\mathrm{i}$, while all remaining eigenvalues have negative real parts. The associated linear frequency is 5.276 Hz, compared with 5.171 Hz on the attracting cycle. Control is applied through $u=\Delta I_{d,\mathrm{ref}}$ with $U=0.15$ p.u. over $T_c=10$ s.
\begin{figure}[!t]
\centering
\subfloat[]{\includegraphics[width=\columnwidth]{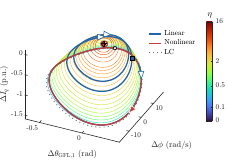}}
\par\vspace{2pt}
\subfloat[]{\includegraphics[width=0.48\columnwidth]{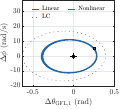}}\hfill
\subfloat[]{\includegraphics[width=0.48\columnwidth]{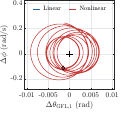}}
\caption{Two-area system: (a) slow-manifold level sets and controlled physical trajectories; (b) trajectories during the final 2 s; (c) enlarged equilibrium neighborhood. Color indicates $\eta$. The circle marks the initial state, the square and diamond mark the linear and nonlinear endpoints, and the cross marks the equilibrium. }
\label{fig:twogeometry}
\end{figure}
The controlled trajectories are compared with the reconstructed manifold in Fig.~\ref{fig:twogeometry}. Under the nonlinear design, the physical trajectory moves inward from the limit cycle and reaches a small neighborhood of the equilibrium. In contrast, the linear design initially reduces the oscillation but leaves a larger residual motion, as shown by the final trajectories in Fig.~\ref{fig:twogeometry}(b)--(c).
\begin{figure}[!t]
\centering
\subfloat[]{\includegraphics[width=0.48\columnwidth]{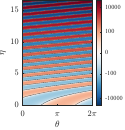}}\hfill
\subfloat[]{\includegraphics[width=0.48\columnwidth]{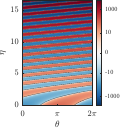}}
\par\vspace{2pt}
\subfloat[]{\includegraphics[width=\columnwidth]{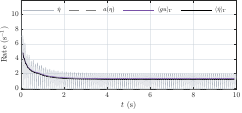}}
\caption{Two-area input responses: (a) $z$; (b) $g$; (c) radial-rate decomposition along the nonlinear design trajectory. Heatmaps use signed symmetric-logarithmic scales; thin contours mark zero. Black curves show one-cycle segments near the beginning, middle, and end. Brackets indicate averages over one cycle period.}
\label{fig:tworesponses}
\end{figure}
This difference can be interpreted through the input responses in Fig.~\ref{fig:tworesponses}(a)--(b). Both $z$ and $g$ vary with phase and radial position, so the allowable input magnitude and its contribution to inward motion change during damping. These variations are retained by the nonlinear design. In Fig.~\ref{fig:tworesponses}(c), the positive cycle-averaged control contribution exceeds the magnitude of the negative autonomous drift, producing net inward progress even when the instantaneous radial rate changes sign.
\begin{figure}[!t]
\centering
\subfloat[]{\includegraphics[width=0.48\columnwidth]{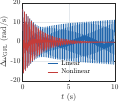}}\hfill
\subfloat[]{\includegraphics[width=0.48\columnwidth]{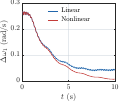}}
\par\vspace{2pt}
\subfloat[]{\includegraphics[width=0.48\columnwidth]{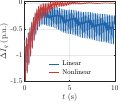}}\hfill
\subfloat[]{\includegraphics[width=0.48\columnwidth]{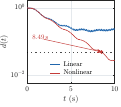}}
\caption{Two-area physical responses: (a) GFL frequency; (b) generator-1 frequency; (c) GFL quadrature-current deviation; (d) normalized state distance, with the 5\% criterion shown by the dotted line.}
\label{fig:twophysical}
\end{figure}
The corresponding full-order responses in Fig.~\ref{fig:twophysical} show that the inward motion is accompanied by reduced inverter-frequency and quadrature-current oscillations, together with a smaller generator-frequency deviation. As reported in Table~\ref{tab:tworesults}, the nonlinear design first enters the 5\% neighborhood at 8.49 s and reaches $d(T_c)=0.02793$. The linear design with $\gamma=1$ reaches a minimum distance of 0.19572 but ends at 0.24786, indicating that the closest approach to the equilibrium is not maintained through the end of the control interval. Increasing the linear gain to $\gamma=2$ improves the terminal distance to 0.21917, which remains above the 5\% threshold. Thus, assigning a stable modal pair in the unsaturated linear model does not ensure comparable suppression of the finite-amplitude physical oscillation.
\begin{table}[t]
\caption{Two-Area Physical-System Results ($T_c=10$ s)}\label{tab:tworesults}
\centering\small\setlength{\tabcolsep}{5pt}
\begin{tabular}{@{}lcccc@{}}\toprule
Design & $d(T_c)$ & $\min d$ & $t_{5\%}$ (s) & $E_u$\\\midrule
Linear, $\gamma=0.5$ & 0.36297 & 0.24312 & -- & 0.02838\\
Linear, $\gamma=1$ & 0.24786 & 0.19572 & -- & 0.03760\\
Linear, $\gamma=2$ & 0.21917 & 0.17719 & -- & 0.04453\\
Slow manifold, $c_\theta=0.4$ & 0.02793 & 0.02793 & 8.49 & 0.03309\\
\bottomrule\end{tabular}
\vspace{2pt}\parbox{\columnwidth}{\footnotesize -- denotes no entry within the control interval. $E_u$ is in p.u.$^2$ s.}
\end{table}
\begin{figure}[!t]
\centering
\subfloat[]{\includegraphics[width=\columnwidth]{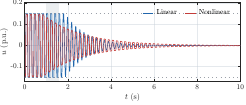}}
\par\vspace{2pt}
\subfloat[]{\includegraphics[width=0.48\columnwidth]{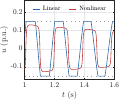}}\hfill
\subfloat[]{\includegraphics[width=0.48\columnwidth]{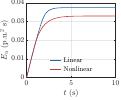}}
\caption{Two-area control inputs: (a) current-reference waveforms; (b) enlargement of the shaded interval; (c) cumulative squared-input effort.}
\label{fig:twoinputs}
\end{figure}
The input comparison in Fig.~\ref{fig:twoinputs} further shows that an 88.7\% reduction in terminal distance relative to $\gamma=1$ is achieved by the nonlinear design while using 12.0\% less squared-input effort. The improvement is therefore obtained within the same actuator limit and with a smaller integrated control input.
\subsection{Modified IEEE 39-bus system}\label{sec:39}
The second study evaluates the same two-state design in a system with detailed synchronous-machine dynamics and multiple GFL inverters. Synchronous generators are retained at buses 30, 31, 32, 37, and 39, while GFL inverters are installed at buses 33, 34, 35, 36, and 38. The synchronous-machine, exciter, and governor models follow \cite{Huang2026TNPF}; the inverter units are represented by the GFL model in \cite{Huang2025SSO}, and the inverter gains are listed in Table~\ref{tab:config}. The equilibrium has one unstable pair $0.0601\pm31.3524\mathrm{i}$, with all other eigenvalues in the open left half-plane. Its linear frequency is 4.990 Hz, whereas the attracting cycle oscillates at 4.635 Hz, showing a larger difference than in the two-area system. Only GFL34 is controlled, through its direct-axis current reference with $U=0.3$ p.u. over $T_c=20$ s. The phase parameter is varied over $c_\theta\in\{0.4,0.5,0.6\}$, with $c_\theta=0.4$ used for the waveform comparisons. 

\begin{figure}[!t]
\centering
\subfloat[]{\includegraphics[width=\columnwidth]{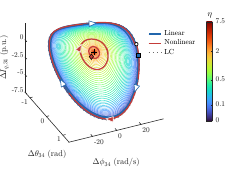}}
\par\vspace{2pt}
\subfloat[]{\includegraphics[width=0.48\columnwidth]{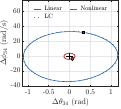}}\hfill
\subfloat[]{\includegraphics[width=0.48\columnwidth]{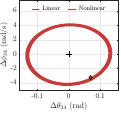}}
\caption{IEEE 39-bus system: (a) slow-manifold level sets and controlled physical trajectories; (b) trajectories during the final 2 s; (c) enlarged equilibrium neighborhood. Color and markers follow Fig.~\ref{fig:twogeometry}.}
\label{fig:39geometry}
\end{figure}

Fig.~\ref{fig:39geometry} shows the manifold and physical trajectories in the controlled inverter's coordinates $(\Delta\theta_{34},\Delta\phi_{34},\Delta I_{q,34})$. The nonlinear design produces a pronounced contraction toward the equilibrium, whereas the trajectory under the linear design remains close to the uncontrolled cycle. The final-interval projections distinguish the small residual oscillation under nonlinear control from the much larger motion under the linear design.

\begin{figure}[!t]
\centering
\subfloat[]{\includegraphics[width=0.48\columnwidth]{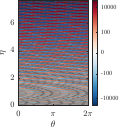}}\hfill
\subfloat[]{\includegraphics[width=0.48\columnwidth]{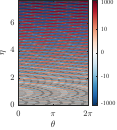}}
\par\vspace{2pt}
\subfloat[]{\includegraphics[width=\columnwidth]{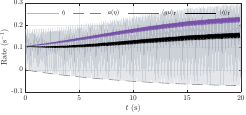}}
\caption{IEEE 39-bus input responses: (a) $z$; (b) $g$; (c) radial-rate decomposition along the nonlinear design trajectory. Plot settings follow Fig.~\ref{fig:tworesponses}.}
\label{fig:39responses}
\end{figure}

The response functions in Fig.~\ref{fig:39responses}(a)--(b) retain the changing phase and radial sensitivities over this transition. Their effect on the control is illustrated in Fig.~\ref{fig:39responses}(c): the cycle-averaged input contribution offsets the outward autonomous drift and maintains positive net radial progress along the reduced trajectory.

\begin{figure}[!t]
\centering
\subfloat[]{\includegraphics[width=0.48\columnwidth]{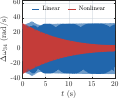}}\hfill
\subfloat[]{\includegraphics[width=0.48\columnwidth]{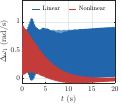}}
\par\vspace{2pt}
\subfloat[]{\includegraphics[width=0.48\columnwidth]{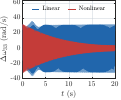}}\hfill
\subfloat[]{\includegraphics[width=0.48\columnwidth]{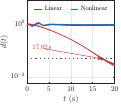}}
\caption{IEEE 39-bus physical responses: (a) controlled GFL34 frequency; (b) generator-1 frequency; (c) unactuated GFL33 frequency; (d) normalized state distance. The dotted line denotes the 5\% criterion.}
\label{fig:39physical}
\end{figure}

The full-order waveforms in Fig.~\ref{fig:39physical} confirm that damping is obtained beyond the controlled inverter. Oscillations at the unactuated GFL33 and generator 1 are also reduced. For $c_\theta=0.4$, Table~\ref{tab:39results} gives a first-entry time of 17.02 s and a terminal distance of 0.02887, compared with 0.97979 for the linear design. This corresponds to a 97.1\% reduction in terminal distance, although only one of the five inverters receives supplementary control.

\begin{table}[t]
\caption{IEEE 39-Bus Physical-System Results ($T_c=20$ s)}\label{tab:39results}
\centering\small\setlength{\tabcolsep}{5pt}
\begin{tabular}{@{}lcccc@{}}\toprule
Design & $d(T_c)$ & $\min d$ & $t_{5\%}$ (s) & $E_u$\\\midrule
Linear, $\gamma=1$ & 0.97979 & 0.82263 & -- & 0.26342\\
Slow manifold, $c_\theta=0.4$ & 0.02887 & 0.02789 & 17.02 & 0.08618\\
Slow manifold, $c_\theta=0.5$ & 0.03177 & 0.02609 & 16.49 & 0.10361\\
Slow manifold, $c_\theta=0.6$ & 0.03460 & 0.02840 & 16.09 & 0.12082\\
\bottomrule\end{tabular}
\vspace{2pt}\parbox{\columnwidth}{\footnotesize -- denotes no entry within the control interval. $E_u$ is in p.u.$^2$ s.}
\end{table}

\begin{figure}[!t]
\centering
\subfloat[]{\includegraphics[width=\columnwidth]{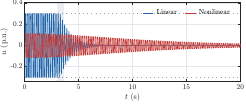}}
\par\vspace{2pt}
\subfloat[]{\includegraphics[width=0.48\columnwidth]{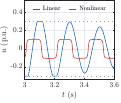}}\hfill
\subfloat[]{\includegraphics[width=0.48\columnwidth]{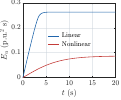}}
\caption{IEEE 39-bus control inputs: (a) current-reference waveforms; (b) enlargement of the shaded interval; (c) cumulative squared-input effort.}
\label{fig:39inputs}
\end{figure}

The control waveforms in Fig.~\ref{fig:39inputs} explain why this improvement cannot be attributed to a larger input. The nonlinear input reaches a peak of 0.1141 p.u., below the common 0.3-p.u. limit reached by the linear input. Its squared-input effort is also 67.3\% lower. Thus, the nonlinear response functions support more effective damping of the physical oscillation with both a smaller peak input and lower integrated effort in this comparison.

\subsection{Residual Oscillations and Parameter Dependence}\label{sec:sensitivity}
To quantify the residual oscillations shown in the waveforms, Table~\ref{tab:rms} compares frequency deviations from equilibrium over the final 2 s. The reductions at the controlled inverters are 96.0\% in the two-area system and 87.3\% in the IEEE 39-bus system. The reductions at generator 1 and the unactuated GFL33 show that the improvement extends to devices without supplementary control.

\begin{table}[t]
\caption{Frequency Deviation During the Final 2 s (rad/s)}\label{tab:rms}
\centering\small\setlength{\tabcolsep}{4pt}
\begin{tabular}{@{}lrrr@{}}\toprule
System / device & Linear & Nonlinear & Reduction (\%)\\\midrule
Two-area, GFL & 7.8087 & 0.3159 & 96.0\\
Two-area, SG1 & 0.0435 & 0.0115 & 73.6\\
IEEE 39, GFL34 & 23.2840 & 2.9520 & 87.3\\
IEEE 39, SG1 & 0.5521 & 0.0279 & 95.0\\
IEEE 39, GFL33 & 22.1671 & 2.8159 & 87.3\\
\bottomrule\end{tabular}
\end{table}

\begin{figure}[!ht]
\centering
\subfloat[]{\includegraphics[width=0.8\columnwidth]{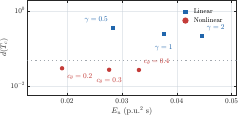}}
\par
\subfloat[]{\includegraphics[width=0.8\columnwidth]{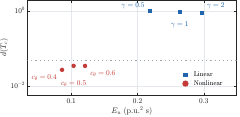}}
\caption{Physical terminal distance versus input effort: (a) two-area system; (b) IEEE 39-bus system. Squares denote linear designs and circles denote slow-manifold designs. }
\label{fig:tradeoff}
\end{figure}

The dependence on control parameters is examined in Fig.~\ref{fig:tradeoff}. The linear gain is varied over $\gamma\in\{0.5,1,2\}$ in both systems, while $c_\theta$ is varied over $\{0.2,0.3,0.4\}$ in the two-area system and $\{0.4,0.5,0.6\}$ in the IEEE 39-bus system. Here $\gamma$ specifies the unsaturated linear modal decay rate, whereas $c_\theta$ limits the phase-speed change under nonlinear control. The initial state, actuator bound, and control horizon are held fixed within each system.

In the two-area system, reducing $c_\theta$ from 0.4 to 0.2 lowers the effort from 0.03309 to 0.01907 p.u.$^2$ s, while the terminal distance increases from 0.02793 to 0.03099. All three nonlinear settings nevertheless remain below the 5\% threshold. In the IEEE 39-bus system, increasing $c_\theta$ produces earlier first entry into the 5\% neighborhood, but also increases the control effort and does not reduce the terminal distance, as shown in Table~\ref{tab:39results}. The setting $c_\theta=0.4$ gives the lowest effort and terminal distance among the three tested values and is therefore used for the principal comparisons. Across the tested settings, the nonlinear designs reach terminal distances below 5\% in both systems, whereas the linear designs do not. These results demonstrate the benefit of retaining phase- and amplitude-dependent input responses when damping sustained nonlinear oscillations.

%% file: sections/appendix.tex
\section{Periodic Coordinate Hessians}\label{app:hessian}
The terminal corrections require the phase and isostable Hessians on $\Gamma$. Let $r=0$ for phase and $r=\kappa$ for isostable, and denote the corresponding gradient and Hessian by $\mathbf{a}_r$ and $\mathbf{H}_r$. Differentiating the coordinate transport identity twice gives
\begin{equation}\label{eq:hequation}
 \dot{\mathbf{H}}_r=r\mathbf{H}_r-\J_\Gamma^{\mathrm T}\mathbf{H}_r
 -\mathbf{H}_r\J_\Gamma-\mathbf{S}_r,
\end{equation}
where $\mathbf{S}_r=\sum_{j=1}^{n}a_{r,j}\nabla^2F_j(\x_\Gamma)$. Symmetry, periodicity, and the tangent constraint $\mathbf{H}_r\F=(r\mathbf{I}_n-\J_\Gamma^{\mathrm T})\mathbf{a}_r$ are imposed jointly. Using the variational matrix $\mathbf{P}$ from Section~\ref{sec:numerical}, integration of \eqref{eq:hequation} with the integrating factors $e^{-rt}\mathbf{P}^{\mathrm T}$ and $\mathbf{P}$ yields
\begin{equation}\label{eq:Hboundary}
 \mathbf{H}_r(0)-e^{-rT_\Gamma}\mathbf{P}_\Gamma^{\mathrm T}
 \mathbf{H}_r(0)\mathbf{P}_\Gamma=\mathbf{C}_r,
\end{equation}
where $\mathbf{P}_\Gamma=\mathbf{P}(T_\Gamma)$ and $\mathbf{C}_r=\int_0^{T_\Gamma}e^{-rt}\mathbf{P}(t)^{\mathrm T}\mathbf{S}_r(t)\mathbf{P}(t)\,\mathrm dt$.

To separate the neutral tangent direction, set $\mathbf{e}=\F(\x_\Gamma(0))/\norm{\F(\x_\Gamma(0))}_2$ and complete it to an orthonormal basis $[\mathbf{e},\mathbf{Q}]$. Since $\mathbf{P}_\Gamma\mathbf{e}=\mathbf{e}$, the remaining monodromy blocks are $\mathbf{a}=\mathbf{Q}^{\mathrm T}\mathbf{P}_\Gamma^{\mathrm T}\mathbf{e}$ and $\mathbf{B}_\perp=\mathbf{Q}^{\mathrm T}\mathbf{P}_\Gamma\mathbf{Q}$. Define $\mathbf{f}_r=(r\mathbf{I}_n-\J_\Gamma(0)^{\mathrm T})\mathbf{a}_r(0)/\norm{\F(\x_\Gamma(0))}_2$. The tangent Hessian blocks are then fixed as $h_{00}=\mathbf{e}^{\mathrm T}\mathbf{f}_r$ and $\mathbf{h}=\mathbf{Q}^{\mathrm T}\mathbf{f}_r$. Consequently, the unknown transverse block $\mathbf{H}_\perp=\mathbf{Q}^{\mathrm T}\mathbf{H}_r(0)\mathbf{Q}$ satisfies
$ \mathbf{H}_\perp-e^{-rT_\Gamma}\mathbf{B}_\perp^{\mathrm T}
 \mathbf{H}_\perp\mathbf{B}_\perp=\mathbf{C}_\perp$
where
$ \mathbf{C}_\perp={}\mathbf{Q}^{\mathrm T}\mathbf{C}_r\mathbf{Q}
 +e^{-rT_\Gamma}\bigl(h_{00}\mathbf{a}\mathbf{a}^{\mathrm T}
 +\mathbf{a}\mathbf{h}^{\mathrm T}\mathbf{B}_\perp
 +\mathbf{B}_\perp^{\mathrm T}\mathbf{h}\mathbf{a}^{\mathrm T}\bigr).
$ 
Thus, only a discrete Lyapunov equation of dimension $n-1$ must be solved. The initial Hessian is reconstructed as $\mathbf{H}_r(0)=h_{00}\mathbf{e}\mathbf{e}^{\mathrm T}+\mathbf{e}\mathbf{h}^{\mathrm T}\mathbf{Q}^{\mathrm T}+\mathbf{Q}\mathbf{h}\mathbf{e}^{\mathrm T}+\mathbf{Q}\mathbf{H}_\perp\mathbf{Q}^{\mathrm T}$. Integration of \eqref{eq:hequation} then supplies the periodic Hessians used in \eqref{eq:terminalphase} and \eqref{eq:terminalcorrection}.